\documentclass[usenatbib,onecolumn]{mn2e}
\usepackage{graphicx}
\usepackage{amssymb}

\usepackage{subfigure}
\newcommand{\beq}{\begin{equation}}
\newcommand{\eeq}{\end{equation}}
 \def\araa{ARAA}
 \def\apj{ApJ}
 \def\pasp{PASP}

 \def\nar{New Ast. Rev.}

\def\araa{ARAA}

\def\apj{ApJ}

\def\beq{\begin{equation}}
\def\ee{\end{equation}}
\def\lsim{\mathrel{\rlap{\lower4pt\hbox{\hskip1pt$\sim$}}
    \raise1pt\hbox{$<$}}}
\def\gsim{\mathrel{\rlap{\lower4pt\hbox{\hskip1pt$\sim$}}
    \raise1pt\hbox{$>$}}}

\begin{document}
\title[Size of discs formed from wind accretion]
{Size of discs formed by wind accretion in binaries can be  underestimated if the role of wind-driving force is ignored}
\author [Blackman, Carroll et al.]
{Eric G. Blackman$^{1}$\thanks{E-mail: blackman@pas.rochester.edu},
Jonathan J. Carroll-Nellenback$^{1}$\thanks{E-mail: johannjc@pas.rochester.edu},
Adam Frank$^{1}$,
\newauthor
 Martin Huarte-Espinosa$^{1}$,
  Jason Nordhaus$^{1,2,3}$\thanks{NSF Astronomy and Astrophysics Postdoctoral Fellow}
\\ $^{1}$Department of Physics and Astronomy, University of Rochester, Rochester NY, 14618, USA\\
$^{2}$Center for Computational Relativity and Gravitation, Rochester Institute of Technology, Rochester NY, 14623, USA\\
$^{3}$National Technical Institute for the Deaf, Rochester Institute of Technology, Rochester NY, 14623, USA\\}

\date{}
\pagerange{\pageref{firstpage}--\pageref{lastpage}} \pubyear{}
\maketitle
\label{firstpage}
\begin{abstract}
Binary systems consisting of a secondary accreting form a wind-emitting primary are  ubiquitous in astrophysics.  The phenomenology of such Bondi-Hoyle-Lyttleton (BHL) accretors is particularly rich when an accretion disc  forms around the secondary.   The outer radius of such discs is  commonly  estimated from the net angular momentum produced by a density variation of material across the BHL or Bondi accretion cylinder,  the latter is tilted with respect to the direction to the primary due to orbital motion.  
But  this  approach has  ignored the fact that the wind experiences an outward driving  force that  the secondary does not.  In actuality, the accretion stream  falls toward a retarded point in the secondary's orbit as the secondary is  pulled toward the primary relative to the stream. The result is a finite 
separation or ``accretion stream impact parameter" (ASIP) separating the secondary and stream.
When the orbital radius $a_o$ exceeds the BHL radius $r_b$, 
 the  ratio of outer disc radius estimated as the ASIP to  the conventional estimate   $ a_o^{1/2}/r_b^{1/2}>1$.
We therefore predict that discs will  form at larger radii from the secondary than traditional estimates.  This agrees with the  importance of the ASIP  emphasized by Huarte-Espinosa et al.   and the practical consequence  that resolving the initial outer radius of such an accretion disc in numerical simulations can be less demanding than what  earlier estimates would  suggest.
\end{abstract}

\section{Introduction}

Accretion discs in binary systems play a fundamental role in the phenomenology of high energy emission from compact objects (for  reviews see Abramowicz \& Straub 2013; Abramowicz \& Fragile 2013; Frank, King, Raine 2002). The compact objects can be black holes, neutron stars, or white dwarfs and so the diversity of accretion in binary systems  represents evolved states of both low mass and high mass stars.  X-ray binaries and microquasars have long been associated with  accretion onto neutron stars and black holes and cataclysmic variables have been associated with accretion in a binary onto white dwarfs.  And, most recently it has been realized that the prevalence of asymmetric and bipolar planetary nebulae (e.g. Bujarrabal et al. 2001; Balick \& Frank 2002)
might also be associated with binaries (e.g. De Marco \& Soker 2009), 
and the  production of jets from the associated accretion discs  (Reyes-Ruiz \& Lopez 1999;  
Soker \& Rappaport 2000,2001; Blackman et al. 2001; Nordhaus \& Blackman 2006; Witt et al. 2009).

If a binary separation is too large for Roche overflow (e.g. Eggleton 1983; D'Souza et al. 2006) or tidal shredding, a disc may still form around the secondary by the accretion of wind material ejected  by the primary. Modeling this mode of accretion disc requires generalizing the so-called Bondi-Hoyle-Lyttleton (BHL) flows (see  Edgar 2004 for a review) to include the  additional asymmetric effects that arise from  orbital motion between the wind emitting primary  and the secondary  (Shapiro \& Lightman 1976; Wang 1981).  
 
 The 
  original (and simplest) form of the BHL problem occurs when a compact object of mass, $M_2$, moves at a constant supersonic velocity $v_{rel}$
 relative to the ambient material though a homogenous plasma  cloud.  
 The gravitational field of $M_2$ focuses the material located within the  ``Bondi cylinder" of radius
\begin{equation}
  r_b = 2GM_2/v_{rel}^2,
  \label{1}
\end{equation}
that extends past the object 
 forming a downstream wake. Downstream, along the Bondi cylinder axis, a stagnation point separates material accreted onto the object
from material that flows away and escapes   (Bondi
\& Hoyle, 1944).   A conical shock forms downstream around the axis
and along this axis the ``accretion line''  (or "accretion stream") connects the stagnation point and the object.

 In addition to phenomenological applications of this mode
 of accretion in binary systems (e.g. Soker \& Rapporport 2000; Struck et al. 2004; Perets \& Kenyon 2012), there have been  a handful of numerical simulations demonstrating the  mechanism  (Smooth particle Hydrodynamics: Theuns \& Jorissen 1993;
Mastrodemos \& Morris 1998,1999;  Grid based:  
Nagae et  al 2004; Jahanara et al. 2005; 
2-D Adaptive Mesh: de Val-Boro et al. 2009; 3-D Adaptive Mesh: Huarte-Espinosa et al. 2013). 
Note also that  there have been many more simulations of the basic BHL mechanism without 
the consideration of the 
angular momentum \cite{2005A&A...435..397F}.

Huarte-Espinosa et al. (2013),  carried out the highest resolution simulations of BHL wind accretion in a binary system 
  in the  limit that the orbital speed is  less than the wind speed, and that the Bondi radius is  less than the orbital scale.
There remains opportunity  to study a  more comprehensive range of companion masses
and orbital radii but an important issue for both conceptual understanding of disc formation and for practical  considerations in  setting up  simulations is the  minimum scale to resolve.  Huarte-Espinosa et al. (2013)  deemed that the minimum scale  to assess the presence or absence
of disc is  the accretion stream impact parameter (hereafter ASIP) $b$.
 This  is the distance of closest approach of the BHL accretion stream to the secondary.  It is  determined by the displacement of the secondary towards the primary over a Bondi accretion time as measured in an inertial frame where the secondary starts from rest and is accelerated towards the primary.  This appears essentially like free-fall towards the slowly moving primary.  In the limited cases studied, the simulations were consistent with this being the
 relevant minimum scale scale needed.   However, Huarte-Espinosa et al. (2013) did not compare this radius to that of standard estimates of the outer disc radius (e.g. Shapiro and Lightman 1976; Wang et al. 1981) based  on the different physics associated with  accretion from a density gradient across the Bondi cylinder. 

In the present paper, we compare the analytic derivations, physics, and distinct predictions of  these two estimates  of the minimum outer accretion disc radius in the limit that the
orbit speed is much less than the wind speed and the Bondi radius is  much less than then orbit radius.   We find that the ASIP does indeed provide the larger scale of the two under the same conditions and thus
 the associated scalings emerge to dominate those of the previous "standard" estimates.
In section 2, we discuss the derivation of the  estimate for the BHL  disc formation based on a 
density gradient and a tilted Bondi cylinder. In section 3 we derive the
ASIP prediction for the distance of closest approach of the accretion
column and we demonstrate that it provides a less restrictive means for
forming disks around secondaries. Finally, conclusions are presented in section 4. 

\section{Estimating  the  Disc Radius from Accretion in a Density Gradient}
 
In this section we derive and synthesize previous estimates  \citep{1976ApJ...204..555S,2005A&A...435..397F} of the disc radius  around the secondary  
based on the density variation and tilt of the BHL accretion cylinder.
The basic  geometry is shown in Fig \ref{fig1} in the frame of the secondary, assuming  a top view of what would be a counter-clockwise orbit in the lab frame. The figure shows increased density of flow lines in the half of the Bondi cylinder where the density is higher.

\subsection{Velocities of the problem}

Although the dimensions are exaggerated for clarity in our figures, we assume in what follows that the Bondi radius of Eq. (\ref{1}) satisfies $R_2<<r_b << a_o$, where $R_2$ is the physical radius of the secondary and $a_o$ is the
orbital separation.
If we define $r$   as the  radial distance  from  the primary to an arbitrary point in the 
rotating frame of the secondary, the relative velocity between the secondary and wind material located at $r$ is 
given by
\beq
v_{rel}^2(r)=v_w^2 +v_{2}^2(r),
\label{2r}
\eeq
where    $v_w$ is the wind speed, and $v_{2}(r)$ is the contribution to velocity from the orbital motion of the secondary.
At $r=a_o$ we write
\beq
v_{rel}^2(a_o)=v_w^2 +v_{2,c}^2,
\label{2rc}
\eeq
where $v_{2,c}=v_2(a_o)$.
At $r=a_o$ we also define $\tan \alpha \equiv v_{2,c}/v_w$
where (as seen in Fig. \ref{fig1}), $\alpha$ is the angle between the direction to the primary and the direction of the wind. 
  In what follows,  we assume that   $v_w >> v_{2,c}$  so that $\alpha$ is a small angle and will keep terms only  first order in $\alpha$. Since $r_b << a_o$, we also have that  deviations from $r=a_o$ are also small, and can be treated as lowest order corrections.

The explicit expression for $v_{2c}$ is given by
\beq
 v_{2,c}= R_{2}\Omega= R_2\left[G(M_1+ M_2)\over a_o^3\right]^{1/2},
 \label{2}
\eeq
which is the circular orbital speed of $M_2$ at displacement  $R_2>0$ from the center of mass. Here $\Omega$ is the orbital speed and  $M_1$ is the mass of the primary.   
   Defining $R_1$ as the magnitude of the displacement of $M_1$  from the center of mass, we can eliminate $R_2$ from Eq. (\ref{2}) by noting that 
\beq
a_o=R_1+R_2=  R_2(1+M_2/M_1)=  R_2 (1 + Q)/Q,
\label{6}
\eeq
where   $Q=M_1/M_2$. Then from Eq. (\ref{2}),  
\beq 
v_{2,c}={Q \over\sqrt{ Q + 1}}\left[GM_2\over a_o\right]^{1/2}, 
\label{7a}\eeq
where we have used Eqn. (\ref{6}).

Eq. (\ref{2}) defines the relative velocity at $r=a_o$,  but for  $r\ne a_o$, 
Eq. (\ref{2r}) requires the position dependent orbital contribution to
the total relative velocity between the wind and the secondary. 
This is given by 
\beq 
v_{2}(r)= \frac{r-R_1}{R_2} v_{2,c} = \frac{r(1+Q)-a_o}{a_o Q} v_{2,c}, 
\label{7bold}
\eeq
which reduces  to $v_{2,c}$ at $r=a_o$.
We  will now   express $v_2$ in coordinates of the  ``Bondi disc'', which we define as the cross section of the Bondi cylinder that intersects the secondary. To do so, we define a coordinate system where the x-axis is perpendicular to the direction of the primary (out the plane in Fig. \ref{fig1} and though the center of the secondary), and the y-axis is oriented in the `almost radial'  direction (for small $\alpha$).
The Bondi disc is seen  edge-on in  Fig. \ref{fig1} 
and indicated by the thick diagonal line through the secondary. 
We can then write
\beq
r= y \sin \alpha + a_o,
\label{ry}
\eeq
noting that $y$ can be negative.
 Then at  any point within the ``Bondi disc'' 
\beq
v_{2}(x,y)\simeq v_{2}(r=y \sin \alpha + a_o)\simeq  \frac{y \sin \alpha + a_o-R_1}{R_2} v_{2,c} = \left (1+\frac{y \sin \alpha}{R_2} \right ) v_{2,c}=
 \left[1+\frac{y \sin \alpha(1+Q)}{a_o Q} \right ] v_{2,c},
\label{7b}
\eeq
where we have used Eqn 
(\ref{6}) for the last equality. Note that for $y<0$, and $\sin \alpha >0$ this formula gives $v_2 < v_{2,c}$ as expected since  being in the orbital
frame means that closer to the center of mass from the position of the secondary, the tangential  velocity for a fixed angular velocity is smaller.

\subsection{Deriving the condition  for disc formation}

The outer boundary of the accretion disc  around the secondary is expected to  form as long as wind material  supplied within the Bondi radius  has  an angular momentum per unit mass about the secondary   equal to or larger than the
angular momentum per unit mass of material in Keplerian orbit $j_2\equiv(GM_2R_k)^{1/2}$ (where $R_k$ denotes the disc outer radius). 
Also, $R_k$ must exceed $R_p$, the physical radius of the secondary for the disc to form.
These conditions for disc formation can  be summarized as 
\beq
R_p < R_k = {j_a^2 \over GM_2}< r_b.
\label{7}
\eeq
 If we write ${j_a}= {{\dot L}\over {\dot M}_b}$, where ${\dot L}$ is the net rate at which angular momentum is
supplied through the Bondi disc and ${\dot M}_b
=\pi r_b^2 \rho(a_o) v_{rel} 
$ 
is the lowest order rate at which mass is supplied through the Bondi disc (ie the Bondi accretion rate uncorrected for the density gradient) then 
assessing the  condition of Eq. (\ref{7}) requires an expression for ${\dot L}$.

We   first derive ${\dot L}$  as an annotated variation  of that  in \cite{1976ApJ...204..555S}  and then comment on its relation to estimates of \cite{1984MNRAS.211..927S}.
Our coordinate system  is rotated from that of \cite{1976ApJ...204..555S} in order to simplify  the visualization and ease the comparison with the next section.

For the relative velocity  in the $-{\hat {\bf z}}$ direction as in Fig 1., the $ {\hat {\bf x}}$ component of angular momentum in a volume 
around the Bondi disc is given by
\beq
{ L}_x 
 =-\int \int \int \rho (x,y,z,t)y v_{rel}{\rm d}z{\rm d}x {\rm d}y
\label{8ang}
\eeq
The time derivative of $L_x$ through the Bondi disc comes from the arrival of mass into the Bondi disc  from the wind coming initially from the $+\hat{ \bf z}$ axis.
We assume that over the scales of interest of  the initial disk formation the relative speed is divergence free thus the continuity equation gives
 $\partial \rho/\partial t = -{\bf v}\cdot \nabla \rho=v_{rel}\partial \rho/\partial z>0$, so  that
 \beq
{\dot L}_x=
-\int \int  \rho(x,y) v_{rel}^2 y  {\rm d}
x {\rm d}y,
\label{8}
\eeq
where the integral is over coordinates in the Bondi disc plane and we have dropped writing the explicit $t$ dependence as we will not need it in what follows.
From  Fig 1. and Eq. (\ref{8}) note that where $y<0$, ${\dot L}_x >0$ as the density is higher than for $y >0$. 

Carrying out the integral in (\ref{8}) requires specification of the position dependences of $\rho$ and $v_{rel}$.
For the density we use 
\beq
\rho(x,y) = \rho(a_o) +{{\partial }\rho(x,y) \over {\partial }y}\Big|_{y=0} y 
\label{9}
\eeq
and  for the relative velocity 
\beq
v_{rel}(x,y) = v_{rel}(a_o) + {{\partial}v_{rel} (x,y)\over {\partial}y}\Big|_{y=0} y.
\label{10}
\eeq
We now need expressions for  ${\rm d}\rho/{\rm d}y$ and ${\rm d}v_{rel}/{\rm d}y$.

Assuming a   wind outflow mass loss rate
${\dot M}_w= 4 \pi r^2 \rho(r) v_w(r) = $constant,
we have
\beq
y{{\rm d}\rho\over {\rm d}y} =y{{\rm d}\rho\over {\rm d}r}{{\rm d}r\over {\rm d}y} =  -2{y\over r}\rho(r){\sin\alpha} 
= -2{y \over a_o} \rho(a_o) \sin\alpha,
\label{14d}
\eeq
to lowest order in $\sin \alpha$, where we have used Eq. (\ref{ry}).
The contribution from the second term on the right of (\ref{9})  is therefore a correction of first order in $\sin \alpha$ compared  to 
the first term on the right
 so that
\beq
\rho =\rho(a_o)\left(1-{2\over a_o}y \sin\alpha\right).
\label{9a}
\eeq

To obtain the velocity correction in  Eq. (\ref{10})  
we  use Eqs. (\ref{2r}) and (\ref{7b})  
to obtain
\beq
y {dv_{rel}\over dy} =y  {v_{rel}\over v_{2}} {dv_2\over dy}
 \sim{y\over a_o} \left({v_{2,c}\over v_{rel}}\right)^2 {(1+Q)\over Q} v_{rel}{\sin \alpha}
 \simeq {y\over a_o}  {(1+Q) \over Q}v_{rel}\sin^3 \alpha,
 \label{16}
\eeq
where we have used $v_2 \simeq v_{2,c}$ and  $\sin \alpha = v_{2,c}/v_{rel}$  in the last two equalities to 
extract the lowest order contribution in $\sin \alpha$.  
As long as $Q > {\sin \alpha \over 1-\sin \alpha}\simeq \sin \alpha$,   Eq. (\ref{16}) represents a smaller than second order correction in $\sin \alpha$  to $v_{rel}$.
We ignore it, while keeping the first order correction to the density of  (\ref{14d}).
We thus approximate (\ref{10}) by
\beq
v_{rel}(x,y) \simeq v_{rel}(a_o).
\label{10a}
\eeq

Eqs. (\ref{9a}) and (\ref{10a}) allow us to integrate Eq. (\ref{8})
which, when converted to cylindrical coordinates $\phi$ and $R={y\over \sin\phi}$ in the Bondi disc, becomes
\beq
{\dot L}_x 
=- \rho(a_o)v_{rel}^2 \int_0^{2\pi}\int_0^{r_b }\left(R\sin\phi-2{R^2\sin^2\phi \over a_o}\sin \alpha \right)R {\rm d}R{\rm d}\phi.
\label{8a}
\eeq
The first integral on the right vanishes by symmetry so the result is
\beq
{\dot L}_x  
= 2{\rho(a_o)v_{rel}^2 \sin \alpha\over a_o} \int_0^{2\pi}\int_0^{r_b } R^3 \sin^2\phi  {\rm d}R {\rm d}\phi=
 {\pi \rho(a_o)v_{rel}^2 r_b^4 \sin \alpha\over 2 a_o}, 
\label{8b}
\eeq
Dividing by 
$
{\dot M}_b=\pi r_b^2 \rho v_{rel}
$
gives 
\beq
j_a=  { v_{rel} r_b^2 \sin \alpha\over 2 a_o} \simeq {v_{2,c}r_b^2 \over2 a_o}
\label{20}
\eeq
for our assumed limit of small $\sin\alpha $  (i.e. $v_w >> v_{2.c}$).  

From (\ref{7a})  (\ref{7}) and (\ref{20}) the  accretion disc radius is then
given by 
\beq
R_k =
{Q^2 r_b^4 \over 4(1 + Q) a_o^3 }.
\label{21}
\eeq

This scaling  of  $r_b^4/a_o^3$ in Eq. (\ref{21})
 agrees with that used in \cite{2000ApJ...538..241S} but 
 disagrees with that derived in  \cite{1984MNRAS.211..927S} who obtained the scaling $r_b^3/a_o^2$ with a coefficient of order unity.
  The  source of  discrepancy  for the latter can be traced to the fact that  starting with the first equality in (\ref{20}) we can also write the condition
  of Eq. (\ref{7}) as  
\beq
R_k 
={v_{2,c}^2Q^2 r_b^3 \over 2 v_{rel}^2 (1+ Q)^2 a_o^2},
\label{22}
\eeq
where we have used $r_b={ 2GM_2\over v_{rel}^2}$ and $\sin \alpha = {v_{2,c}\over v_{rel}}$.
We therefore see that in (\ref{22}), the scaling of $r_b^3/a_o^2$ and the coefficient that depends on the velocity
ratios would be $O(1)$ only if  $v_{2,c} >> v_w$, so that $v_{rel}\sim v_{2,c}$. This would correspond to the axis of the Bondi cylinder 
almost perpendicular to the orbital radius. This  limit violates our  assumption  
that $\sin \alpha = v_w/v_{2,c}$ is small so the scaling of  \cite{1984MNRAS.211..927S} does not apply.  In fact if we use Eq. \ref{1} and \ref{7a} to eliminate the velocities from Eq. \ref{22} we recover that $R_k \propto r_b^4/a_o^3$.

 In short Eqn. (\ref{22}) is the self-consistent relation.  Livio et al. (1986)  find correction coefficients between $0.1$ and $1$  to this formula that depend on adiabatic index.  Since our present focus is the radial scaling and the ratio of Eq. (\ref{22}) to Eq. (\ref{24}) of the next section,  we ignore 
 those corrections for present purposes.

\section{Including the wind driving force trumps the  estimate of the previous section} \label{sec3}

\begin{figure}
\centering

  \includegraphics[width=\columnwidth]{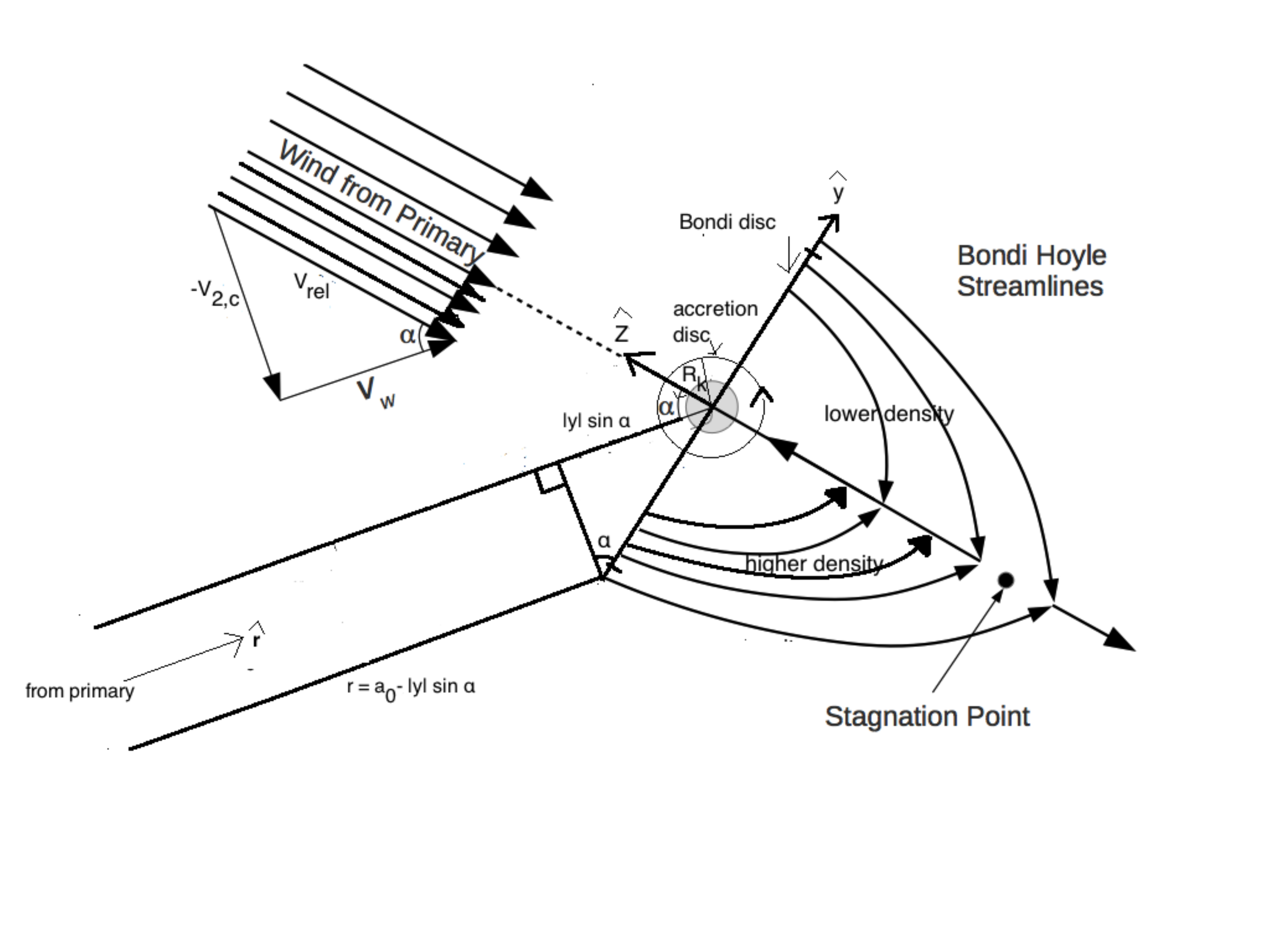}
  \caption{Schematic showing the flow structure of the binary wind capture
  process used in the ``standard" approach to calculate the radius of disc formation.
  The figure is shown in the instantaneous inertial rest frame frame of the secondary. In the lab frame the secondary
  would be orbiting counterclockwise from this top view.  The higher density of lines indicate a higher density of wind material
  because that part of the flow is closer to the primary. The purported radius of the accretion disk forming about the secondary is shown as
  $R_k$ as discussed and computed in section 2. As stated in the text, we assume $sin \alpha<<1$ but we have exaggerated the angle in the figure
  for clarity. The Bondi disc seen edge-on is indicated by the line segment between the two hash marks on the $\hat {\bf y}$ axis. }
  \label{fig1}
\end{figure}

\begin{figure}
\centering
  \includegraphics[width=\columnwidth]{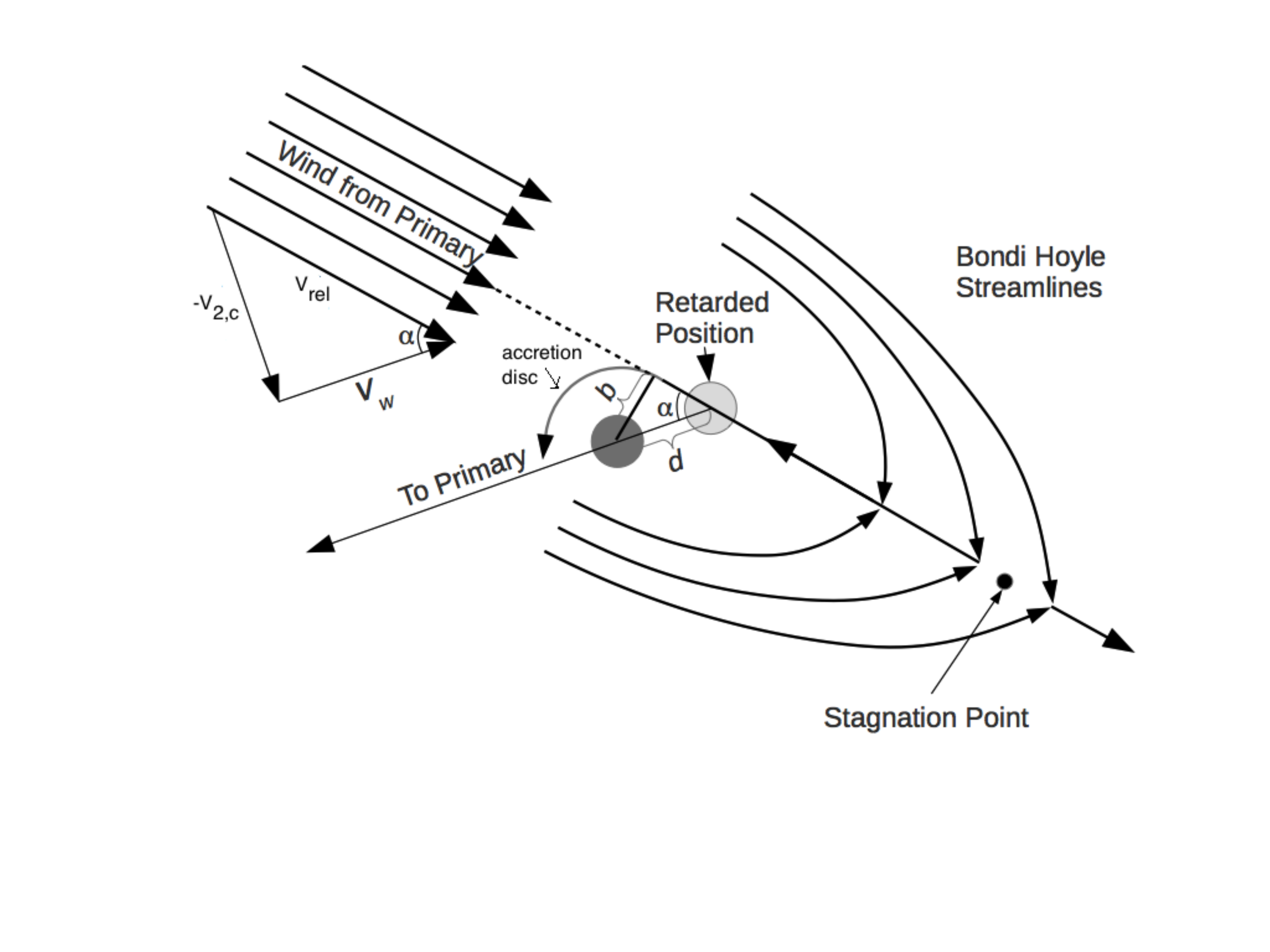}
  \caption{Same as Fig.1 except that here the density variation across the Bondi cylinder is ignored, and instead  the  acceleration of the secondary toward the primary relative to that  of the wind is included. The two positions of the secondary that are shown at different times highlights that this figure is drawn in the inertial frame at the  retarded time not the rotating frame.
  This leads to the accretion stream falling toward a retarded position of the secondary
		 and the formation of an accretion disc with radius $b$ as discussed in the text.
  }
  \label{impactParameter}
\end{figure}

The  calculation of the previous section ignores an important effect.  
The wind is being accelerated by a outward  force (such as radiation pressure)  not felt by the secondary.
Radiation force scales with distance from the primary as  $r^{-2}$ and under the assumed conditions of the previous section, namely $r_b << a_o$, 
such a scaling implies that the wind experiences approximately  the same outward force on time scales for which the secondary makes any displacement of scale   $b\le r_b$ toward the primary in the secondary's inertial frame. 
As a consequence of the secondary's orbital acceleration, gas from the stagnation point
does not flow directly toward the secondary but  toward a retarded position  between the secondary's earlier and current positions.
Fig. ~\ref{fig1}  shows the flow structure in the instantaneous inertial frame of the orbiting secondary.
Fig. ~\ref{impactParameter} is a modification of Fig. \ref{fig1} to remove the effect of density variation,  but to include the retardation effect just described.
In this section we show that even without the density variation the displacement $b$ (the ASIP), exceeds $R_k$ estimated in the previous section and provides a larger predicted accretion disc radius.  Note that  we have assumed that the a balance has been achieved between gravity and the wind driving force such that $V_w$ is constant.  Thus the force does not appear explicitly in our calculations

To calculate the predicted disc scale analogue of (\ref{20})  from the retardation effect,  note that 
the time scale, $t_c$, associated with the wind capture process scales
approximately with the time scale for material to flow from the stagnation point to 
the retarded position of the secondary. It takes approximately a
free-fall time, $t_{ff}$, for material to fall toward the secondary,
and the stagnation point is at least as large as the Bondi radius so a conservative estimate of
$t_c$  can be made by assuming the material  falls a displacement,
$r_b$, giving
\beq 
t_c \simeq  t_{ff} ={\pi\over 2} \frac{r_b^{3/2}}{(2GM_2)^{1/2}} \sim \frac{r_b}{v_{rel}}.
\label{21}
\eeq
During this time, 
the secondary is accelerated out of the instantaneous rest frame towards the primary relative to the wind with 
acceleration $A_c\sim \frac{v_{2,c}^2}{R_2}$.
The distance, $d$, travelled by the secondary during the wind capture
time scale $t_c$ is 
\begin{equation}
   d = \frac{1}{2} A_c t_c^2,
\label{d_sec}
\end{equation}
\noindent and the distance perpendicular to the accretion 
   column
is  the ASIP
\begin{equation}
   b = d \sin{\alpha} = d \frac{v_{2,c}}{v_{rel}},
\label{par2.5}
\end{equation}
using the defintion of $\alpha$.
\noindent 
Then using the expressions for $d$, $A_c$ and $t_c$ we obtain
 \begin{equation}
b 
=\frac{ v_{2,c}^3r_b^{2}(Q+1)}{2v_{rel}^3a_oQ}=\frac{ r_b^{7/2}Q^2}{2^{5/2}a_o^{5/2}(Q+1)^{1/2}},
\label{24}
\end{equation}
where in the first equality we have used Eq. (\ref{6}) to convert from $R_2$ to $a_o$ and in the second equality we have
used  Eqn. (\ref{1}) and 
 (\ref{7a}).

We emphasize that $b$ measures the effective impact parameter of the accretion stream to the secondary, and the accretion stream
moves approximately in free-fall from the stagnation point. Since the stagnation point is located at  a position $\infty > r \gsim r_b$ from the point of closest approach,
 as long as $b<< r_b$, the flow at this position of closest approach would  achieve a speed $v_b$  close to, but 
 just below the escape speed. Specifically  $v_{b} \approx \sqrt{2GM_2/b}$.  This would correspond to a specific angular momentum of $j \approx \sqrt{2 GM_2 b}$ or a circular orbit or radius $\approx 2b$.  

Having established that  material at a displacement  $b$  from the secondary meets the criterion of sufficient angular momentum,
we can now ask which estimate, $b$ or $R_K$ determined in the previous section is larger. That will determine where the disc forms.
The  ratio  of  $b/R_k$   from (\ref{24}) and  (\ref{21}) is 
\beq
{R_k\over b}= \frac{2^{1/2} r_b^{1/2}}{a_o^{1/2}(Q+1)^{1/2}}.
\label{25}
\eeq
This ratio is important because in the present approximation  that $r_b << a_o$,  we see that 
$b > R_K$.  Because $b$ is the minimum scale for a disc to form
around $M_2$ when the effect of wind acceleration is accounted even in the absence of a density gradient,  
 such a  disc will form at a larger radius than estimated by the method of the previous section.
As an aside, note that the direction of angular momentum of the accretion disc is the same for both estimates (compare Figs. 1 and 2).



%
{

%


%
%

\section{Conclusion}

In a binary system where the primary is well within its Roche radius and accretion occurs onto the secondary via the well known BHL
 accretion of the primary's wind, an accretion disc can still form around the secondary if the matter flowing onto the secondary has enough  angular momentum  to exceed the Keplerian speed of an orbit  at its surface.  We have considered the condition for initial disc formation for the  case in which both the orbital separation well exceeds the Bondi accretion radius around the secondary and the orbital speed of the secondary is much less than the  wind speed from the primary.  We have shown that previous standard estimates for the size of the accretion disc, derived from combining the tilt between the binary radial separation vector and the axis of the Bondi accretion cylinder and the associated wind density variation across the Bondi cylinder,  underestimate  the disc radius.  

The underestimate is a consequence of the incorrect  assumption  that the accretion stream follows a trajectory aimed toward the 
current position of the secondary.   Correcting this assumption  means taking into account the fact that the accretion stream is aimed at a retarded position of the secondary because the latter does not feel the accelerating force experienced by the wind.
The secondary is drawn toward the primary by unfettered gravity during a Bondi accretion time.  Equivalently one can consider the torque about the secondary applied by the wind on the accretion stream as it falls towards the secondary. 
 The offset between the present and retarded positions of the secondary defines the ASIP in Eq. (\ref{22}) which provides the larger accretion disc radius 
 estimate as summarized by Eq. (\ref{25}).  
 
For the parameters of Huarte-Espinosa et al. (2013) $v_{2,c}/v_w \sim 0.5$ and $a_o/r_b ~ 1.56$, so the regime is not quite the asymptotic regime
that we have studied herein.  Nevertheless, they were on the right track in emphasizing the potential importance of  Eq. (\ref{22}) 
as the minimum scale needed to  numerically resolve the initial formation of an accretion disc  from wind accretion by an orbiting secondary.
(Once a substantial disc forms however,  a higher resolution may   be required to study the  subsequent  interaction between wind and  disc.)
Natural desired  generalizations of the present work   are   indeed  cases  for which $r_b$ is not that much less than $a_o$ and/or $v_w < v_{2,c}$, and/or the density falls off faster than $1/r^2$.   In such cases, the calculations of  sections 2 and sections 3 should be revised to include higher order corrections.  In such cases, the Bondi cylinder  would be more strongly asymmetric in density so that the actual accretion disc radius would be a combination of the retardation effect and the density variation.
In addition, if  $r_b$ is comparable to $a_o$,  the acceleration force of the wind would vary significantly over the scale of the accretion stream flow from the stagnation point to the secondary. 
Thus the relative acceleration of the secondary toward the primary compared to that of the wind would not be a constant, as we have assumed in this paper. but would spatially vary across the trajectory of the accretion stream.  We leave these generalizations as future opportunities.

\section*{Acknowledgments} 
  We acknowledge  support from NSF Grants PHY-0903797 and AST-1109285.  JN is supported by an NSF Astronomy and Astrophysics Postdoctoral Fellowship under award AST-1102738 and by NASA HST grant AR-12146.01-A. We thank M. Ruffert for comments.

\bibliographystyle{mn2e}

\end{document}